\documentclass[american,aps,prx,reprint,superscriptaddress, longbibliography,floatfix]{revtex4-2}
\usepackage[dvips]{graphicx}
\usepackage{amsmath,amssymb,physics,amsthm,mathrsfs,amsfonts,dsfont}
\usepackage{subcaption, epsfig}
\usepackage{ragged2e}
\usepackage{caption}
\DeclareCaptionJustification{fillspace}{\justifying}

\usepackage{braket}
\usepackage{bm}
\usepackage{enumerate}
\usepackage[svgnames]{xcolor}
\usepackage{qcircuit}
\usepackage{algorithm}
\usepackage{algpseudocode}
\usepackage{hyperref}
\usepackage{orcidlink}
\hypersetup{
     colorlinks = true,
     linkcolor = Maroon,
     anchorcolor = black,
     citecolor = DarkBlue,
     filecolor = blue,
     urlcolor = DarkBlue
     }
\usepackage{comment}
\usepackage{ulem}

\newcommand{\Var}{\mathrm{Var}}

\usepackage{varioref}
\labelformat{equation}{#1}

\labelformat{section}{#1}

\labelformat{figure}{#1}

\labelformat{proposition}{#1}

\labelformat{lemma}{#1}

\labelformat{theorem}{#1}

\labelformat{observation}{#1}

\labelformat{definition}{#1}

\labelformat{problem}{#1}

\labelformat{algorithm}{#1}

\labelformat{corollary}{#1}

\labelformat{statement}{#1}

\begin{document}

\title{Entanglement-facilitated macroscopic cluster formation in \\quantum many-body dynamics
}

\author{Xiao Wang\,\orcidlink{0000-0002-3022-7260}}
\affiliation{School of Physical and Chemical Sciences, Queen Mary University of London, London, E1 4NS, United Kingdom}
\affiliation{
Department of Physics, Tsinghua University, Beijing 100084, China}
\affiliation{Clarendon Laboratory, University of Oxford, Parks Road, Oxford, OX1 3PU, United Kingdom}

\author{Alexander Yosifov}
\affiliation{Clarendon Laboratory, University of Oxford, Parks Road, Oxford, OX1 3PU, United Kingdom}
\affiliation{School of Physical and Chemical Sciences, Queen Mary University of London, London, E1 4NS, United Kingdom}

\author{Aditya Iyer}
\affiliation{Clarendon Laboratory, University of Oxford, Parks Road, Oxford, OX1 3PU, United Kingdom}

\author{Jinzhao Sun}
\email{jinzhao.sun.phys@gmail.com}
\affiliation{School of Physical and Chemical Sciences, Queen Mary University of London, London, E1 4NS, United Kingdom}

\date{\today}

\begin{abstract}

Metastable quantum many-body dynamics could facilitate the organisation of microscopic degrees of freedom into macroscopic structures. However, the conditions under which this occurs are not well understood. Here we study false-vacuum decay in a 2D quantum Ising model and show that the initial correlation structure can qualitatively change this behaviour. Compared with product-state initialisations, correlated false-vacuum states suppress the proliferation of small true-vacuum domains and favour the formation of macroscopic connected clusters. Tree tensor network simulations of lattices up to $25 \times 25$ further reveal that nucleation proceeds predominantly from the boundary. Finite-size scaling demonstrates the dominant connected cluster remains an extensive fraction of the system even as its size increases. By suppressing this edge-assisted nucleation pathway through boundary pinning, we generate large magnetisation fluctuations consistent with macroscopic superposition. This mechanism relies on the 2D nucleation barrier and is absent in 1D systems or product-state quenches. Our results identify correlated state preparation and boundary engineering as complementary techniques for controlling metastable quantum dynamics.
\end{abstract}

\maketitle

The evolution of interacting quantum many-body systems is accompanied by the formation and reorganisation of spatial structures \cite{polkovnikov2011colloquium,eisert2015quantum,gogolin2016equilibration,nahum2017quantum}. Understanding how microscopic dynamics can give rise to macroscopic structures is of broad interest. In condensed matter physics, they reflect emergent collective many-body behaviour, while in quantum information processing, preserving extended domains can provide a route to the distribution of global information or passive protection of information \cite{RevModPhys.87.307,RevModPhys.88.045005,bravyi2009no}. A central open question thereby is: under what conditions does many-body dynamics organise into macroscopic connected structures?

Metastability---the long-lived persistence of an excited configuration---provides a natural framework here \cite{turner1982our, coleman1980gravitational,vodeb2025stirring}, as it captures the competition between relaxation and the survival of system-size structures. A useful perspective is provided by the statistics of connected clusters \cite{saberi2015recent}, where the presence of large components indicates macroscopic connectivity. While in the quantum setting, connected clusters provide a natural measure of spatial correlations \cite{stauffer2018introduction, PhysRevLett.102.100401}. 

Still, the conditions that allow macroscopic clusters to form and persist remain unclear. Existing works on false vacuum (FV) decay~\cite{PhysRevB.104.L201106,chao2026probing,osterholz2025collective,darbha2024false,zenesini2024false,vodeb2025stirring,borla2026microscopic,pavevsic2025scattering,krinitsin2025time,song2022realizing,zhu2024probing} point to two key ingredients. First, dimensionality matters: in 1D local defects fragment extended domains, whereas in 2D a nucleation barrier can support large connected regions \cite{langer1969statistical, binder1987theory, sachdev1999quantum}. Second, most studies of quantum quench dynamics on Rydberg analogue simulators or quantum annealers rely on product states \cite{polkovnikov2011colloquium, berges2004prethermalization,chao2026probing,osterholz2025collective,darbha2024false,vodeb2025stirring,zenesini2024false}, which are experimentally convenient but lack nonlocal correlations and tend to generate local excitations that drive rapid fragmentation \cite{calabrese2006time, rigol2008thermalization, lerose2019impact,chao2026probing}. Yet, the interplay of the two has not been systematically studied partly due to the difficulty of simulating higher-dimensional dynamics by classical methods or preparing correlated states in analogue quantum simulators. As a result, the importance of preparing correlated initial states in higher-dimensional systems to control the formation and persistence of system-size connected domains remains largely underexplored.

Here, we focus on how the initial correlation structure affects the formation and persistence of macroscopic connected clusters, and how these dynamics can be engineered toward macroscopic quantum superposition (i.e., cat state). As a testbed, we consider the 2D transverse longitudinal field Ising model (TLFIM) as the minimal model which allows us to study how nucleation dynamics, quantum correlations, and boundary engineering jointly control the stability of extended structures. We study the non-equilibrium dynamics following a sudden inversion of the longitudinal field, which places the system in a metastable configuration. We compare the evolution of product states with that of correlated initial states via matrix product state (MPS) and tree tensor network (TTN) methods up to size of $25\times 25$~\cite{verstraete2008matrix,schollwock2011density,PhysRevLett.93.076401,lagnese2021false,10.21468/SciPostPhys.15.4.152,pavevsic2025scattering,krinitsin2025time}. We show that initial-state entanglement qualitatively alters the resulting cluster statistics: compared with product states, correlated states suppress the proliferation of small domains and favour larger connected clusters. Interestingly, finite-size scaling shows the dominant cluster remains macroscopic as the system size increases. Further, we show that pinning the boundary suppresses edge-assisted nucleation, providing a way to enhance macroscopic fluctuations without imposing periodic boundary conditions.

Unlike prior works that mainly focus on FV decay rates or global observables, we study the spatial organisation of the relaxation process. By resolving the evolution of true-vacuum (TV) bubbles via connected-cluster statistics, we show the initial correlation structure determines not only how quickly the system relaxes, but also the decay pathways. This spatial framework enables the identification and control of dominant nucleation channels, and opens new avenues for programmable quantum simulators.

\begin{figure}
    \centering
    \includegraphics[width=0.95\linewidth]{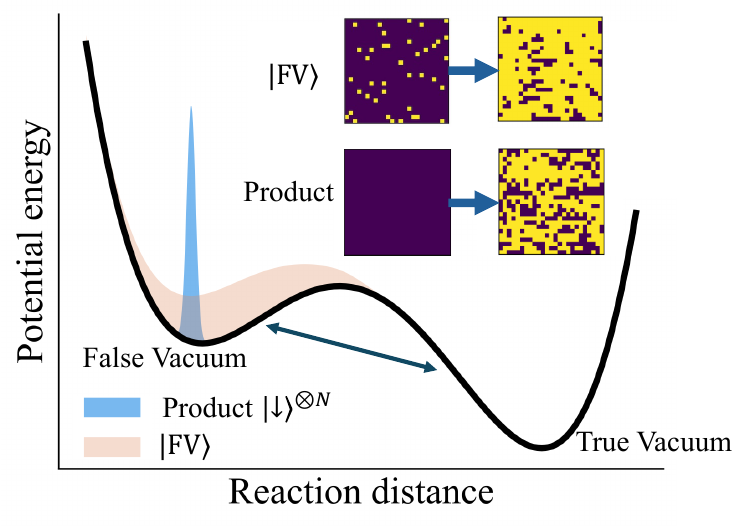}
    \caption{{Metastable energy landscape and cluster dynamics.} The energy landscape of FV decay in the studied 2D Ising model, where the horizontal axis is the Hamming distance relative to the TV state, while the vertical line is the probability density. The $\ket{\psi_{0}}=\bigotimes_{i}\ket{\downarrow}_{i}$ state is a delta function with a fixed distance from the TV. In contrast, $\ket{\text{FV}}$ is a superposition of configurations with varying Hamming distances. Projective spin snapshots on a $L=25$ lattice of the time evolution from $t=0$ to $t=25$ are also shown. After a quench with Eq. (\ref{eq:Ising}), the state starting from $\ket{\rm FV}$ maintains large connected clusters (yellow).}
    \label{fig:fig1}
\end{figure}

\begin{figure*}[ht!]
    \centering
\includegraphics[width=1.0\linewidth]{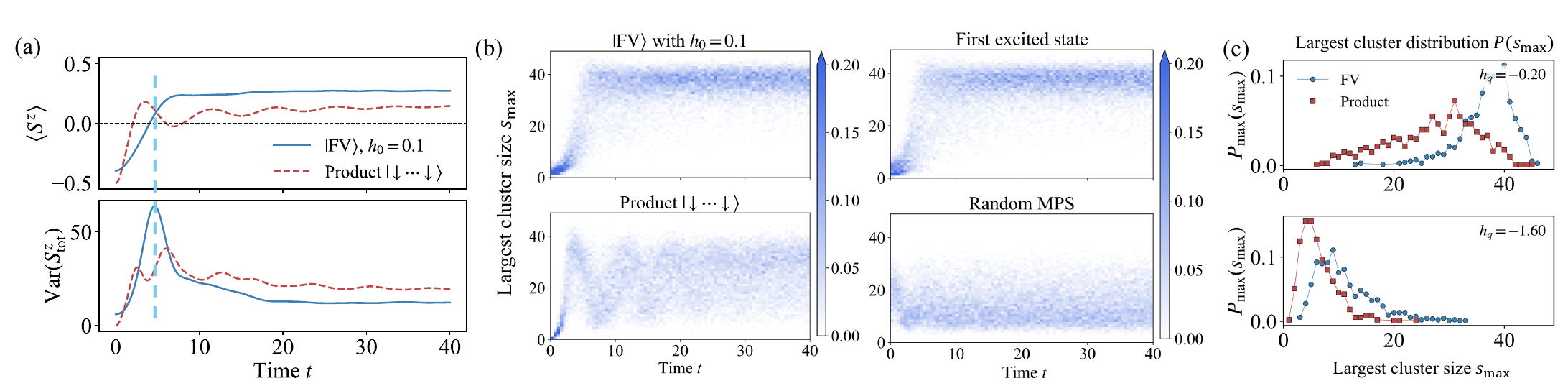}
    \caption{
    {Quench dynamics of the 2D Ising model on an \(L=7\) lattice simulated using MPS.} $\ket{\text{FV}}$ is prepared with a small symmetry-breaking field $h_0=0.1$; the bond dimension is set to $\chi=256$, while $J=-1$, $g=1$, and $h_q=-0.2$. (a) Time evolution of the average magnetisation \(\langle S^z \rangle\) and the variance of the total longitudinal magnetisation \(\Var( S^z_{\mathrm{tot}})\).
    (b) At each time, the largest cluster size \(s_{\max}\) is extracted from \(800\) projective \(S^z\)-basis snapshots, and the resulting distribution \(P_{\max}(s_{\max},t)\) is shown as a colormap. \textbf{Left:} Evolution starting from $|\mathrm{FV}\rangle$ (top), and from $\ket{\psi_0}$ (bottom). \textbf{Right:} Evolution starting from the first excited state of the pre-quench Hamiltonian (top), and from a random MPS state with entanglement entropy comparable to that of $|\mathrm{FV}\rangle$ (bottom). The colour scale is capped at \(P_{\max}=0.2\), with larger values saturated. (c) Percolation-oriented cluster observables extracted from $800$ projective $S^z$-basis measurements of the post-quench state at $t=40$ for $h_{q}\in\{-0.2, -1.6\}$. The \(h_q=-0.2\) data correspond to a final-time cut of the two left panels in (b). The largest-cluster distribution $P_{\text{max}}(s_{\max},t)$ is defined as the probability that the largest connected flipped cluster in a measurement shot has size $s_{\max}$ at $t=40$. A spin is counted as flipped relative to $\ket{\psi_{0}}$, and the connected clusters are defined using nearest-neighbour connectivity. 
    } 
    \label{fig:max_cluster_size_dynamics}
\end{figure*}

\textit{Setup and metastability.---}We consider the 2D TLFIM on an open $L\times L$ square lattice with spin-$\tfrac{1}{2}$ degrees of freedom as the minimal model for studying information spreading.
The Hamiltonian is given by \cite{sachdev1999quantum}
\begin{equation}
\label{eq:Ising}
H=J\sum_{\langle ij\rangle} S_i^z S_j^z + g\sum_i S_i^x + h\sum_i S_i^z ,
\end{equation}
where $J=-1$ is the ferromagnetic Ising coupling, $S_i^p$, $p \in \{x,y,z \}$ are the spin-$\tfrac12$ operators acting on site $i$, $g$ is the transverse field, and $h$ is the longitudinal field.
The TLFIM provides a minimal setting where \(J\) generates ordered domains, $g$ gives the quantum fluctuations necessary to build entanglement in the ground state (before quench), and $h$ induces symmetry-breaking toward one ferromagnetic domain. We set $g=1$, for which the system retains a two-sector ferromagnetic ground-state landscape. We prepare the FV as the ground state for a small symmetry-breaking field \(h_0=0.1\), and then quench the longitudinal field to \(h_q<0\), making the initial state metastable with respect to the post-quench Hamiltonian. Its subsequent decay is driven by bubble nucleation.

Here, the transition from FV to TV is via the nucleation and subsequent growth of TV bubbles, rather than through a simultaneous reversal of the entire system \cite{turner1982our, coleman1980gravitational}. We can examine the role of the initial state by mapping the many-body system's state on the energy landscape, defined by the Hamming distance with respect to the TV state, which serves as a global coordinate that tracks the metastable decay, Fig.~\ref{fig:fig1}. As detailed below, the initial state preparation dictates how the system occupies this landscape. 
We distinguish two classes of initial states: (i) all-down product state $\ket{\psi_{0}}=\bigotimes_{i}\ket{\downarrow}_{i}$ and (ii) correlated FV state $\ket{\text{FV}}$. $\ket{\psi_{0}}$ represents a singular point in the Hamming landscape and is characterised by a delta-function distribution at fixed distance from the TV. Physically, this implies the state lacks the fluctuations necessary to probe the surrounding configuration space at $t=0$. Consequently, the decay proceeds via the stochastic creation of local excitations, leading to the rapid fragmentation of the global-sized domain as the seeds proliferate. On the other hand, $\ket{\text{FV}}$ has a finite width in its Hamming distribution, representing a coherent superposition of configurations with varying distances from the TV. This spectral broadening indicates the state already incorporates the structured entanglement and domain-wall fluctuations inherent to 2D.

This enables distinct pathways between Hamming sectors. In 2D, a transition in the global Hamming distance requires the formation of adjacent clusters to be energetically favourable. For $|\psi_0\rangle$, these clusters are generated stochastically, whereas the initial spread of $|\text{FV}\rangle$ across the landscape suggests the correlated state already samples the barrier region through domain-wall fluctuations. Thereby, $|\text{FV}\rangle$ reconfigures the tunnelling and activation pathways, shifting the dynamics from stochastic fragmentation to a globally coherent evolution.

\textit{Real-time dynamics.---}To investigate the role of initial correlations, we simulate the non-equilibrium evolution following a quench into the metastable regime. We quantify the relaxation through the average magnetisation
$  
\langle S_z\rangle(t)=\frac{1}{L^2}\sum_i\langle S_i^z\rangle=\frac{1}{L^2}S^z_{\text{tot}},
$ 
which measures the persistence of the initial order in the lattice. 

In Fig.~\ref{fig:max_cluster_size_dynamics}(a) we start with the behaviour of \(\langle S^z \rangle\), where both states begin from a FV polarisation, but $\ket{\psi_0}$ shows a sharper initial growth indicative of swift decay via uncorrelated local spins that act as nucleation seeds for TV bubbles \footnote{Ref.~\cite{pavevsic2025scattering} considers the TLFIM in a weaker-field regime, $|g/J|\simeq0.75$ and $|h_q/J|\lesssim0.15$ in our spin-half convention, where intrinsic false-vacuum decay is strongly suppressed on the simulated timescale; in that regime, decay is instead triggered by injecting external energy through wave-packet collisions.}. Interestingly, we also observe larger transient build-up of longitudinal magnetisation fluctuations \(\Var( S^z_{\mathrm{tot}})=\langle (S^z_{\mathrm{tot}})^2 \rangle - \langle S^z_{\mathrm{tot}} \rangle^2\) by $\ket{\text{FV}}$. For pure states, this is directly proportional to the quantum Fisher information with respect to $S^z_{\mathrm{tot}}$. 
This early spike suggests dynamical generation of macroscopic correlations, indicating that initial-state entanglement shifts the decay mechanism from uncorrelated fragmentation to macroscopic correlated fluctuations. The degree to which this depends on lattice geometry is examined in SM Sec.\@\ref{sec:SMsecA}.

\textit{Investigation of the cluster statistics from metastable states.---}A relevant question here is what are the underlying spatial structures that protect the FV from decay in 2D? To determine how stability emerges, we examine the time evolution of the largest-cluster distribution $P_{\max}(s_{\max},t)$. Here $s_{\max}$ is the largest flipped cluster (defined by the horizontally or vertically connected spin-up sites) in a $Z$-basis projection measurement snapshot.

For $\ket{\text{FV}}$ in Fig. \ref{fig:max_cluster_size_dynamics}(b), it is clear that the probability weight rapidly concentrates at large $s_{\max}\sim 35-40$, marking the formation of a system-size flipped domain. In contrast, $\ket{\psi_0}$ exhibits a broader, fragmented distribution with pronounced intermediate oscillations, indicating reduced macroscopic connectivity. To further isolate the role of the initial state, we compare these results to the first excited state and a random MPS state with entanglement entropy comparable to that of $\ket{\text{FV}}$. Here we observe that the excited state replicates the rapid concentration at large $s_{\max}$, similar to the FV case. Notably, the clusters generated from a random MPS state \cite{page1993average} remain broadly distributed at small and intermediate sizes. This observation implies that the formation of large clusters is not merely a consequence of entanglement entropy alone, but rather depends on the specific pre-quench correlation structure that suppresses bubble proliferation and sustains macroscopic connectivity. Meanwhile, $P_{\max}(s_{\max},t)$ in Fig. \ref{fig:max_cluster_size_dynamics}(c) further supports this picture. Compared with $\ket{\psi_0}$, $\ket{\mathrm{FV}}$ exhibits substantially higher weight at large $s_{\max}$, indicating the formation of macroscopic connected domains and the emergence of global connectivity.

\begin{figure*}[ht!]
    \centering
\includegraphics[width=1.0\linewidth]{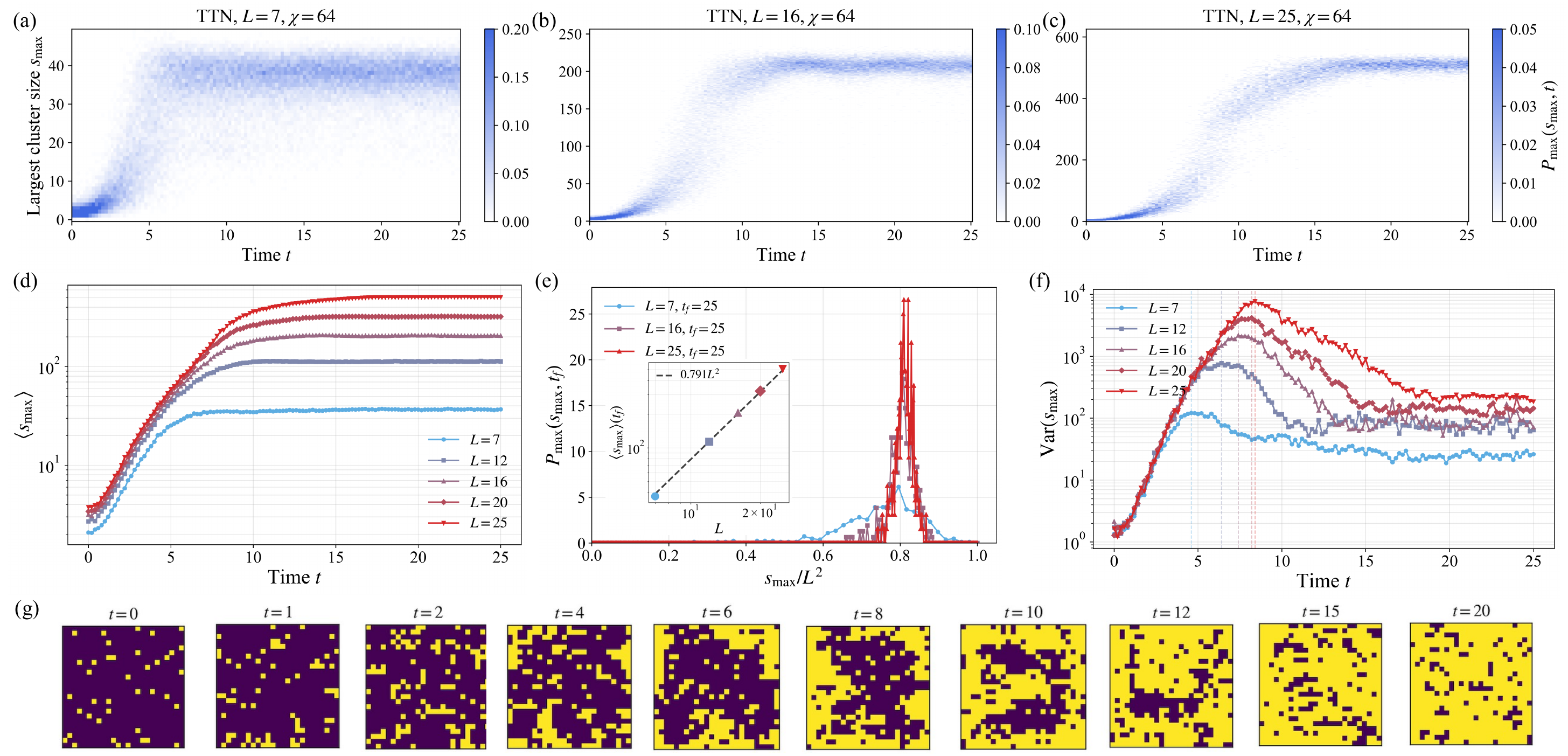}
    \caption{{Dynamics and emergence of macroscopic connected clusters.} (a,b,c) Dynamics of the largest-cluster-size distribution \(P_{\max}(s_{\max},t)\) for $\ket{\text{FV}}$ on $L=7,16,25$ lattices, respectively, simulated using a TTN with bond dimension $\chi=64$.   
    (d) Time evolution of the mean largest-cluster size $\langle s_{\rm max}\rangle$. 
    (e) Probability distribution of the largest flipped-spin cluster at \(t_f=25\), plotted against the scaled cluster size $s_{\max}/L^2$ for $L=7,16,25$. Inset: the scaling of $\braket{s_{\max}}$ over $L$ at $t_f$, which is about $\langle s_{\max} \rangle = 0.79 L^{2}$. 
    (f) Time evolution of the variance of the largest-cluster size $\Var(s_{\max})$. (g) Projection measurement snapshots from $t=0$ to $t=20$ showing the decay of $\ket{\text{FV}}$. Evidently, flipped domains first aggregate near the boundary and subsequently merge into a connected macroscopic cluster.
    }
    \label{fig:TTN_large_cluster_dynamics}
\end{figure*}

Physically, these observations underscore the stabilising role of the initial correlation structure in FV decay. After the quench, the product state $\ket{\psi_0}$ predominantly generates uncorrelated local spin-flips, leading to the rapid nucleation and growth of TV bubbles that destabilise the FV. The associated rapid return-probability decay is analysed in SM Sec.\@\ref{sec:SMsecA1}. In contrast, the pre-quench ground state encodes coherent superpositions of domain-wall configurations that correlate local relaxation processes and suppress dephasing, thereby extending the lifetime of the initial order \cite{kormos2017real}. As a result, flipped regions grow and merge more collectively, favouring the formation of a system-size connected cluster. The contrasting intermediate-time distributions \(P_{\max}(s_{\max},t)\) in Fig.~\ref{fig:max_cluster_size_dynamics}(b) therefore indicate qualitatively different decay pathways, rather than merely delayed relaxation (see SM Sec.\@\ref{sec:SMb1} for additional simulations).

\textit{Macroscopic cluster formation.---}In order to answer whether this dynamical cluster formation is macroscopic, we need to investigate how connected-cluster statistics $s_{\max}$ scale with $L$. However, $L=7$ simulations already approach the classical limits of MPS-based methods for large $t$. To address this, we apply TTNs which allow us to access much larger systems (see SM Sec.\@\ref{sec:SMb2}).

First, in Fig.~\ref{fig:TTN_large_cluster_dynamics}(a,b,c) we show the dynamics of $s_{\max}/L^2$ for $L=7,16,25$ with $\ket{\text{FV}}$, in which the convergence of TTN is verified by increasing the bond dimension to $\chi = 64$.
Then, to see the macroscopic nature of the cluster, we show the scaling behaviour in Fig.~\ref{fig:TTN_large_cluster_dynamics}(e); it shows the final-time distribution of the largest connected flipped-spin cluster, shown as a function of the scaled cluster size $s_{\max}/L^2$ for $L=7,16,25$. The collapse of the distributions near $s_{\max}/L^2\simeq 0.79$ indicates that, at late times, the dynamics is dominated by a macroscopic cluster occupying a fixed fraction of the 2D system.  

We study the variance of the largest cluster size $s_{\max}$ throughout the evolution in Fig.~\ref{fig:TTN_large_cluster_dynamics}(f). By the time $\Var(s_{\max})$ achieves the largest value, we find correspondingly in Fig.~\ref{fig:TTN_large_cluster_dynamics}(a,b,c) the distribution $P_{\text{max}}(s_{\max},t)$ forms a gap, representing the leap of the size of the largest cluster; this is consistent with percolation-like connectivity as the domain grows.
We also find that $\Var(s_{\max})$ at earlier times is nearly identical for different $L$.

\begin{figure}[t!]
    \centering
\includegraphics[width=1.0\linewidth]{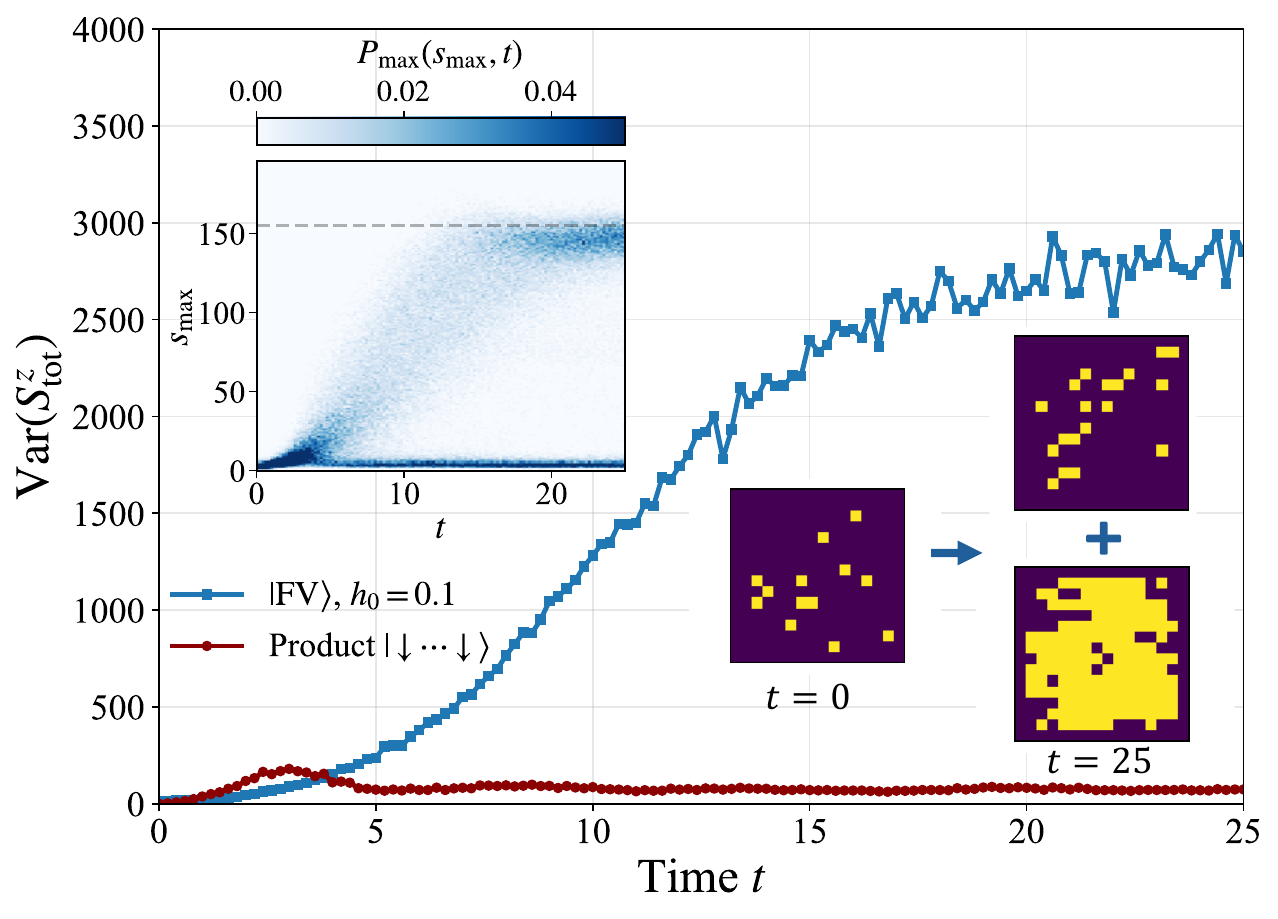}
    \caption{{Boundary pinning enhances macroscopic magnetisation fluctuations.} $\mathrm{Var}(S^z_{\rm tot})$ is extracted from $800$ projective $S^z$-basis measurements at each time. A strong longitudinal edge field $h_{\rm edge}=20$ pins the boundary spins toward the $\ket{\downarrow}$ configuration, suppressing edge-assisted nucleation. For $\ket{\text{FV}}$, edge pinning produces a large growth of $\mathrm{Var}(S^z_{\rm tot})$, whereas the dynamics from $\ket{\psi_0}$ remains weakly fluctuating. Inset: largest-cluster statistics for $L=16$ edge-pinned dynamics. The active bulk has length $L-2$, and the dashed line marks $0.79(L-2)^2$, obtained from the scaling in Fig.~\ref{fig:TTN_large_cluster_dynamics}(e). 
    Although the initial FV is nearly uniform, the edge-pinned unitary dynamics generates large fluctuations associated with a two-branch largest-cluster distribution.
    }
    \label{fig:TTN_edge_var}
\end{figure}

\textit{Enhancing macroscopic fluctuations via edge engineering.---}Evolution snapshots for an $L=25$ lattice are shown in Fig.~\ref{fig:TTN_large_cluster_dynamics}(g), indicating that, for open boundaries, TV domains preferentially nucleate near the edge, where the domain-wall cost is reduced. Motivated by this observation, we suppress boundary-assisted nucleation by pinning the edge spins with a strong longitudinal field $h_{\rm edge}=20$ on the boundary sites, favouring $\ket{\downarrow}$ at the edges. This corresponds to adding the boundary term $h_{\rm edge}\sum_{i\in \text{edge}} S_i^z$ to Eq. (\ref{eq:Ising}) and defining $\ket{\text{FV}}$ as the ground state of the resulting pre-quench Hamiltonian. Starting from $\ket{\text{FV}}$, prepared with the bulk longitudinal field $h_0=0.1$ in the presence of the edge-pinning field, we quench the bulk field to $h=-0.2$ while keeping $h_{\rm edge}$ fixed.

As shown in Fig.~\ref{fig:TTN_edge_var}, edge pinning strongly enhances $\Var(S^z_{\rm tot})$ for dynamics starting from $\ket{\rm{FV}}$, while the $\ket{\psi_0}$ dynamics remain close to the unpinned case. The inset shows the largest-cluster distribution develops both high- and low-\(s_{\max}\) branches, indicating a coherent splitting between an FV-survival branch and a decay branch containing a macroscopic connected flipped cluster, i.e., a macroscopic cat state. The growth of $\Var(S^z_{\rm tot})$ shows the evolved state acquires substantial weight in macroscopically distinct magnetisation sectors. Thus, boundary pinning suppresses an edge-nucleation channel and amplifies the macroscopic fluctuations generated from $\ket{\text{FV}}$.

Why is the cat state generated in the edge-pinned 2D $L\times L$ model absent in a pinned 1D chain with the same total number of $L^2$ spins? Edge pinning suppresses the preferred boundary-nucleation channel of the FV, playing a role analogous to using periodic boundaries. This effect is much stronger in 2D: an $L\times L$ lattice has $O(L)$ pinned boundary sites, whereas a 1D chain has only two pinned endpoints. However, suppressing boundary nucleation alone is not sufficient to generate a macroscopic cat state. 
The essential additional ingredient is 2D connectivity. Under a snake-like ordering, the 2D lattice can be viewed as a stretched chain composed of $L$ length-$L$ segments. If these segments evolved independently, their contributions to $S^z_{\mathrm{tot}}$ would add incoherently and, according to the central limit theorem, produce an approximately Gaussian distribution, as observed in a genuine $L^2$-site 1D chain~\footnote{The corresponding segments in a genuine 1D chain are not completely independent, since neighbouring segments are connected end to end. However, these correlations are weak compared with the 2D case, where transverse bonds directly couple sites across different segments.}. In the stretched 2D representation, by contrast, the vertical bonds become inter-segment Ising couplings. These couplings correlate nucleation across different segments, causing large flipped domains to appear or disappear collectively (2D) rather than independently (1D). Consequently, the distribution of $S^z_{\mathrm{tot}}$ becomes strongly non-Gaussian and develops two macroscopically separated peaks.

Edge pinning surpresses the dominant boundary nucleation, while 2D connectivity makes the remaining bulk dynamics collective. Their combination allows the wavefunction to split into two macroscopically distinct branches: one remains close to the FV, while the other contains a large flipped cluster. This produces both the branching of $s_{\max}$ and the large variance of $S^z_{\mathrm{tot}}$. In a pinned 1D chain, the absence of transverse inter-segment couplings prevents the same non-Gaussian two-branch structure from emerging. See SM Sec.\@\ref{sec:SMb2} for numerical comparisons between 1D and 2D cases.

\textit{Discussion.---}Our work shows how the initial correlation structure qualitatively reshapes FV nucleation in 2D. Although the post-quench spectrum governs the unitary evolution \cite{PhysRevLett.110.135704}, the spatial organisation of the decay is controlled by both lattice dimensionality and the many-body structure of the initial state. Correlated $\ket{\rm{FV}}$ suppress fragmentation and favour macroscopic connected clusters, while large-scale TTN simulations reveal edge-assisted nucleation as the dominant decay channel. Suppressing this channel via boundary engineering enhances macroscopic fluctuations and yields two macroscopic branches. Thus, metastable decay is controlled not only by nucleation energetics, but also by initial correlations and boundary geometry.

Our results establish connected-cluster statistics as a complementary framework for studying quantum metastability. Unlike return probabilities, which depend primarily on the spectral distribution of the initial state \cite{PhysRevLett.96.140604}, cluster observables directly resolve the spatial organisation and distinct pathways of the decay. More generally, the results show the spatial structure in quantum metastability can be a controllable resource. The identification of edge-assisted nucleation suggests boundary engineering as a practical route for controlling metastable dynamics. In programmable quantum simulators, such as Rydberg atom arrays \cite{chao2026probing,browaeys2020many}, correlated state preparation together with engineered boundaries could suppress dominant decay channels and enhance macroscopic fluctuations without active error correction. Combining this passive stabilisation with feedback control or active correction may further sustain or grow macroscopic connected clusters.

Finally, the resulting connected-cluster distributions define a physically motivated sampling problem generated by interacting quantum dynamics. Digital quantum computers can access these distributions through projective measurements, suggesting cluster statistics as a possible benchmark for quantum simulators and a new class of many-body quantum sampling tasks.

\textit{Acknowledgments.---}X.W. and A.Y. contributed equally to the theoretical development of this work. We would like to thank Masanori Hanada for the useful discussions. The \texttt{QTea} TN package \cite{qtealeaves} was used for the numerical simulations. A.Y. and A.I. are supported by UKRI Future Leaders Fellowship (Grant No. 10128920). This research acknowledges funding from the UK EPSRC through EP/Z53318X/1 and support from Schmidt Sciences LLC.

\bibliography{references.bib}

\clearpage
\widetext

\section*{Supplemental Materials}
\phantomsection
\label{sec:SM}

\subsection{Dimensionality of the lattice}
\label{sec:SMsecA}
\subsubsection{Geometry dependence of return probability and magnetisation dynamics}
\label{sec:SMsecA1}
\begin{figure}[ht!]
    \centering
    \includegraphics[width=0.45\linewidth]{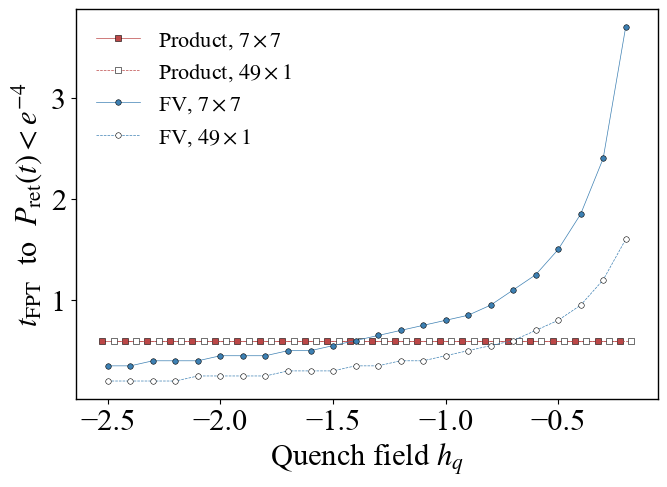}
    \caption{{First-passage time $t_{\mathrm{FPT}}$ for the return probability to fall below the threshold $P_{\mathrm{ret}}(t)\le e^{-4}$}. We compare the $L=7$ and $49\times 1$ lattices for $h_q\in[-2.5, -0.5]$, $J=-1$, $g=1$, and $h_0=0.1$. As $h_q$ approaches zero from below, $\ket{\text{FV}}$ shows pronounced increase of $t_{\mathrm{FPT}}$, particularly in the $L=7$ geometry, signalling enhanced metastability. In contrast, the results for $\ket{\psi_0}$ are nearly identical for the two geometries.}
    \label{fig:quench_dynamics_L7}
\end{figure}

\textit{Return probability and first-passage time.---}To examine the lattice-geometry dependence discussed in the main text, we compare 1D and 2D dynamics using the first-passage time \(t_{\mathrm{FPT}}\), defined as the time at which $P_{\mathrm{ret}}(t)\le e^{-4}$, corresponding to decay in the non-perturbative regime, Fig. \ref{fig:quench_dynamics_L7}. Here, $P_{\mathrm{ret}}(t)$ is the return probability \cite{PhysRevA.30.1610}
\begin{equation}
P_{\mathrm{ret}}(t)=|\langle\psi(0)|\psi(t)\rangle|^2,
\end{equation}
which probes the coherence loss relative to the initial state \cite{gorin2006dynamics}. We see that the dynamics starting from the all-spin-down product state $\ket{\psi_0}$ is nearly geometry-independent, which follows the mean-field estimate for the all-down configuration \cite{PhysRevLett.101.120603}
\begin{equation}
P_{\mathrm{ret}}(t)\approx \exp(-Ng^{2}t^{2}/4).  
\end{equation}
In contrast, $\ket{\rm{FV}}$ depends sensitively on both the dimensionality and $h_{q}$. Strikingly, notice that for strong $h_{q}$, $\ket{\text{FV}}$ initially exhibits shorter $t_{\text{FPT}}$ than $\ket{\psi_0}$, suggesting initial correlations may actually accelerate the FV decay. As $\vert h_{q} \vert$ decreases, however, $t_{\text{FPT}}$ for $\ket{\text{FV}}$ grows sharply and eventually crosses the baseline of $\ket{\psi_0}$, entering a regime of enhanced metastability. As expected, this crossover occurs at smaller values of $h_{q}$ for the $L=7$ lattice compared to the $49\times1$ geometry. The qualitative difference between the 1D and 2D geometries comes from the existence of a nucleation barrier in the latter. In 2D, the competition between the domain-wall energy and the bulk energy gain produces a critical bubble size $R_c$, whereas in 1D the energy decreases monotonically with $\ell$ once a kink-antikink pair has formed. Consequently, metastable decay in 2D proceeds through the nucleation and subsequent growth of $R>R_c$ bubbles.

Furthermore, the existence of $R_c$ provides a useful interpretation of the metastable landscape in Fig.~\ref{fig:fig1}(a). Although the Hamming distance is not identical to the bubble radius $R$, it serves as a collective coordinate describing the progress of the decay. Configurations close to the metastable minimum correspond predominantly to small, $R<R_c$ bubbles, whereas increasing Hamming distance reflects the growth of TV domains. Crossing the nucleation barrier therefore corresponds to the onset of irreversible bubble expansion.

The dimensionality divergence in Fig. \ref{fig:quench_dynamics_L7} highlights the inability of mean-field methods to describe entangled states. Recent works on simulating FV decay mostly focus on initial product states, which are restricted by the controllability of Rydberg analogue simulators \cite{chao2026probing,osterholz2025collective,darbha2024false}, quantum annealers~\cite{vodeb2025stirring} or the capability of TNs \cite{borla2026microscopic,pavevsic2025scattering}. As shown here, the decay of correlated $\ket{\rm{FV}}$ is governed not only by nucleation energetics, but also by the underlying structure of many-body quantum correlations, which qualitatively alters the fragmentation dynamics.  The detectable features, such as $P_{\text{ret}}(t)$, will thus behave qualitatively differently from entangled states in various dimensions.

\textit{BCH expansion analysis.---}Following the Baker-Campbell-Hausdorff analysis of Ref.~\cite{chao2026probing}, we consider the early-time dynamics of the site-averaged magnetisation
\[
\langle S^z\rangle(t)
\equiv
\frac{1}{N}\sum_i\langle S_i^z\rangle_t
=
\frac{1}{N}\langle S^z_{\mathrm{tot}}\rangle_t,
\]
where \(N=L^2\) for the square lattice. 
Both initial states can be represented by real wavefunctions in the \(S^z\) basis, up to an irrelevant global phase. Since the post-quench Hamiltonian \(H\) and total magnetisation \(S^z_{\mathrm{tot}}\) are real symmetric in this basis, the odd-order nested commutators \(\operatorname{ad}_{H}^{2m+1}(S^z_{\mathrm{tot}})\) are real antisymmetric and therefore have vanishing expectation values in either initial state. Hence only even powers of \(t\) appear, giving
\begin{equation}
    \langle S^z\rangle(t)
    =
    \langle S^z\rangle(0)
    -\frac{t^2}{2}C_2
    +O(t^4).
\end{equation}
The second-order coefficients are
\begin{equation}
    C_2^{(\psi_0)}
    =
    -\frac{g^2}{2},
    \qquad
    C_2^{(\mathrm{FV})}
    =
    -\frac{g(h_q-h_0)}{N}
    \langle S^x_{\mathrm{tot}}\rangle_{\mathrm{FV}},
\end{equation}
while for $\ket{\psi_{0}}$
\begin{equation}
    \langle S^z\rangle_{\psi_0}(t)
    =
    -\frac12+\frac{g^2t^2}{4}+O(t^4),
\end{equation}
so its leading short-time curvature is independent of the post-quench longitudinal field $h_q$. In contrast, the curvature for $\ket{\rm{FV}}$ is linear in the quench amplitude \(h_q-h_0\). The correlated $\ket{\rm{FV}}$ therefore exhibits a substantially stronger dependence on the longitudinal-field quench already at order \(t^2\), whereas the product-state response at this order is controlled solely by the transverse field. This short-time distinction is consistent with the contrasting \(h_q\)-dependence of the first-passage times shown in Fig.~\ref{fig:quench_dynamics_L7}.

\textit{Spectral invariance and spatial observables.---}Our results clearly show that 2D initially entangled states retain global clusters. While the presence of long-lived plateaus in the return probability $P_{\text{ret}}(t) = |\langle \psi | e^{-i \hat{H}t} | \psi \rangle|^2$ suggests a high degree of global coherence within the metastable regime, it is essential to distinguish between the stability of this global overlap and the physical stability of the state's structure. Because $P_{\text{ret}}(t)$ depends solely on the energy spectral distribution of the initial state \cite{PhysRevLett.110.135704,PhysRevLett.96.140604}, that limits its utility as a probe of metastability. By expanding the initial state $|\psi\rangle = \sum_n c_n |E_n\rangle$ in the eigenbasis of the Hamiltonian $\hat{H}$, the amplitude is given by $A(t) = \sum_n |c_n|^2 e^{-iE_n t}$. It is thus easy to see that any unitary transformation $\hat{U}$ that commutes with the Hamiltonian leaves $P_{\text{ret}}(t)$ invariant, as such operators preserve the spectral weights $|c_n|^2$. A direct consequence of this symmetry is that the time-evolved state $|\psi(t')\rangle = e^{-i \hat{H}t'}|\psi\rangle$ shares the same return probability as the initial state for any $t'$, despite the fact that local observables and entanglement entropy may have evolved substantially by time $t'$ \cite{lieb1972finite}. This spectral invariance highlights a key limitation: $P_{\text{ret}}(t)$ probes only how much the evolved and initial states overlap, but not how correlations are spatially arranged in the lattice. This underscores the need for percolation-based cluster analysis \cite{cardy1992critical} to faithfully characterise the spatial organisation of metastable dynamics.

\subsubsection{Energetic considerations: surface-volume competition in 1D and 2D bubbles}

The results above are consistent with an energetic distinction between 1D and 2D FV decay. In 2D, the energy of a TV bubble is controlled by the competition between a positive domain-wall contribution and a negative bulk-energy gain. On a lattice, the precise coefficients depend on the bubble geometry and microscopic details
\[
E_{\mathrm{2D}}(R)\simeq A\sigma R-B\Delta\epsilon R^2,
\]
where \(A\) and \(B\) are model-dependent geometrical factors, \(\sigma\) is the domain-wall tension, and \(\Delta\epsilon>0\) is the energy-density difference between the TV and FV~\cite{pavevsic2025scattering}. This surface-volume competition yields a critical bubble size \(R_c\sim\sigma/\Delta\epsilon\): bubbles with \(R<R_c\) tend to shrink, whereas those with \(R>R_c\) expand. This energetic structure favours collective, compact flipped domains rather than many independently nucleated small segments. At late times, neighbouring flipped clusters can merge into a system-size domain, after which the isolated-bubble picture is replaced by one of residual unflipped FV islands embedded in a TV background. In this late-time regime, further growth of the flipped TV domain can \textit{reduce} the total boundary energy, in contrast to the isolated-bubble picture.

In contrast, a 1D bubble is an interval of length \(\ell\) bounded by two domain walls, with energy
\[
E_{\mathrm{1D}}(\ell)\simeq 2\sigma-\ell\Delta\epsilon,
\]
which decreases monotonically with \(\ell\) once the kink-antikink pair has formed. Thus, 1D systems do not possess an analogous critical bubble size \(R_c\). This is consistent with the picture discussed above: in a genuine long 1D chain, different length-\(L\) segments are only weakly correlated, so their contributions to \(S^z_{\rm tot}\) tend to add incoherently rather than forming two sharply separated macroscopic branches. In 2D, by contrast, the transverse bonds correlate the nucleation dynamics of different segments and favour collective flipped domains. Consequently, the extensive connected clusters and the large fluctuations of \(S^z_{\rm tot}\) observed in Figs.~\ref{fig:TTN_large_cluster_dynamics} and~\ref{fig:TTN_edge_var} are not expected to arise in the same way in 1D.

\subsection{Tensor network simulations}
\subsubsection{Supplemental simulations for Fig. 2}
\label{sec:SMb1}
\begin{figure*}[ht!]
    \centering
    \includegraphics[width=0.85\textwidth]{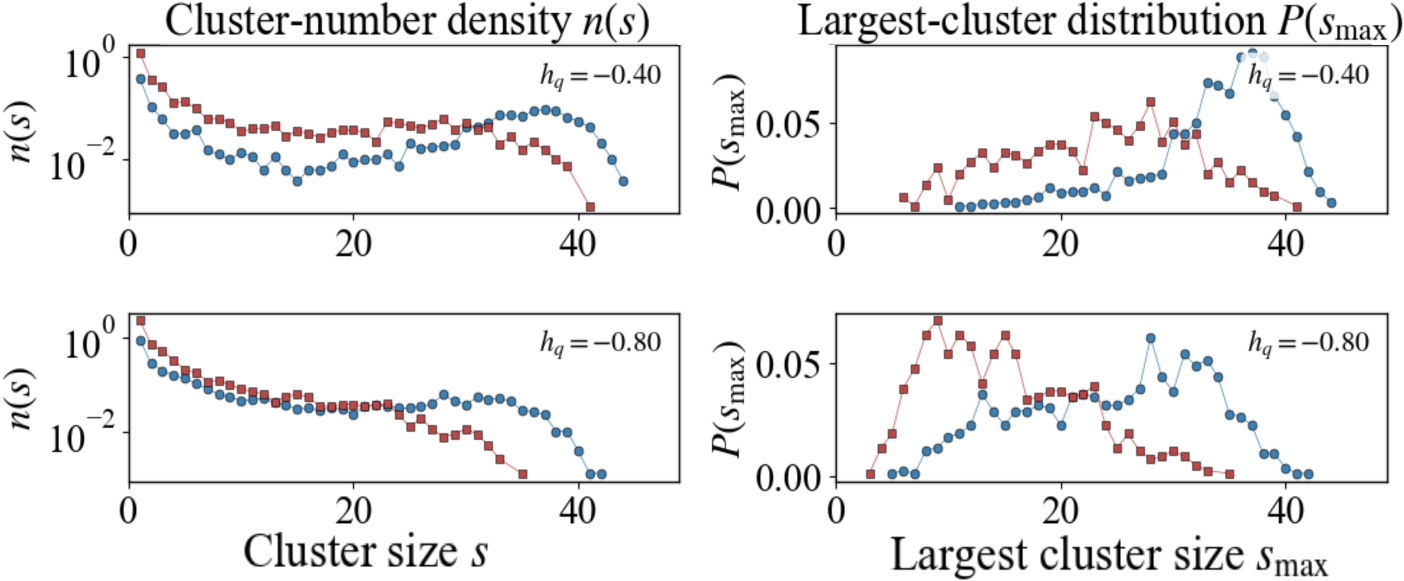}
    \caption{Extension of the percolation-oriented cluster observables results shown in Fig.~\ref{fig:max_cluster_size_dynamics} in the main text.}
    \label{fig:cluster_statistics}
\end{figure*}

Here we provide additional simulations supporting the results in Fig.~\ref{fig:max_cluster_size_dynamics} of the main text, including additional quench strengths and cluster observables, finite-size comparisons, and dynamics from different pre-quench eigenstates.
Fig.~\ref{fig:cluster_statistics} extends the cluster observables shown in Fig.~\ref{fig:max_cluster_size_dynamics} of the main text to different values of $h_q$ and additionally presents the cluster-number density $n(s)$. 
For both values of $h_q$, the largest-cluster distribution for $\ket{\text{FV}}$ is shifted toward larger cluster sizes relative to that for $\ket{\psi_0}$, indicating a higher probability of system-size clusters. 
Similarly, $n(s)$ exhibits a higher density of large clusters for $\ket{\rm{FV}}$, including a characteristic increase at large $s$, demonstrating that the system sustains macroscopic connected domains even under the stronger quench.

\begin{figure*}[ht!]
    \centering

    \begin{subfigure}[t]{0.48\linewidth}
        \centering
        \caption{}
        \includegraphics[width=\linewidth]{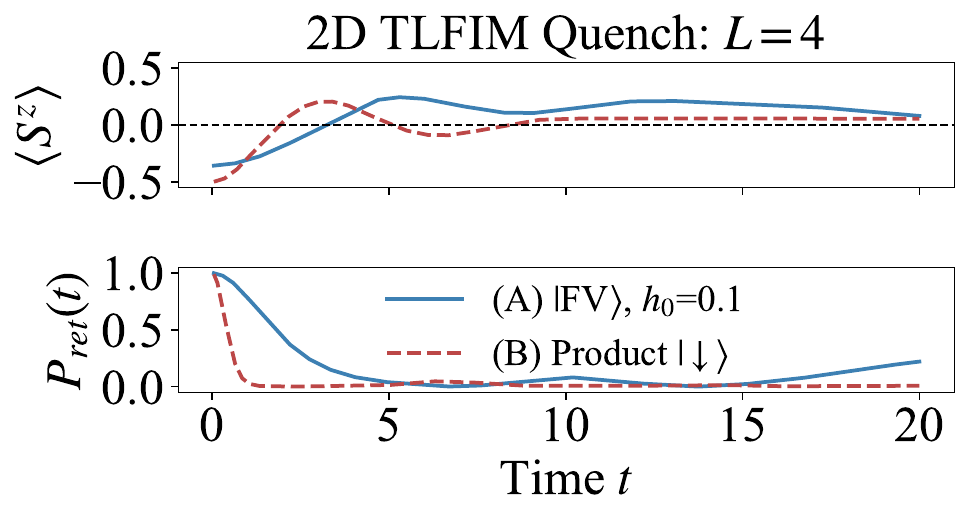}
        \label{fig:tlfim_quench_4by4}
    \end{subfigure}
    \begin{subfigure}[t]{0.48\linewidth}
        \centering
        \caption{}
        \includegraphics[width=\linewidth]{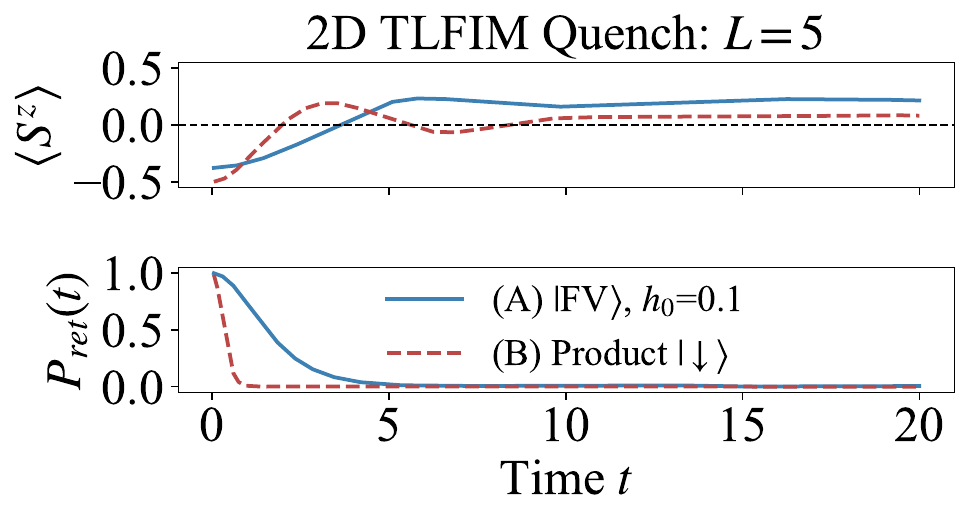}
        \label{fig:tlfim_quench_5by5}
    \end{subfigure}

    \vspace{-0.8em}

    \begin{subfigure}[t]{0.48\linewidth}
        \centering
        \caption{}
        \includegraphics[width=\linewidth]{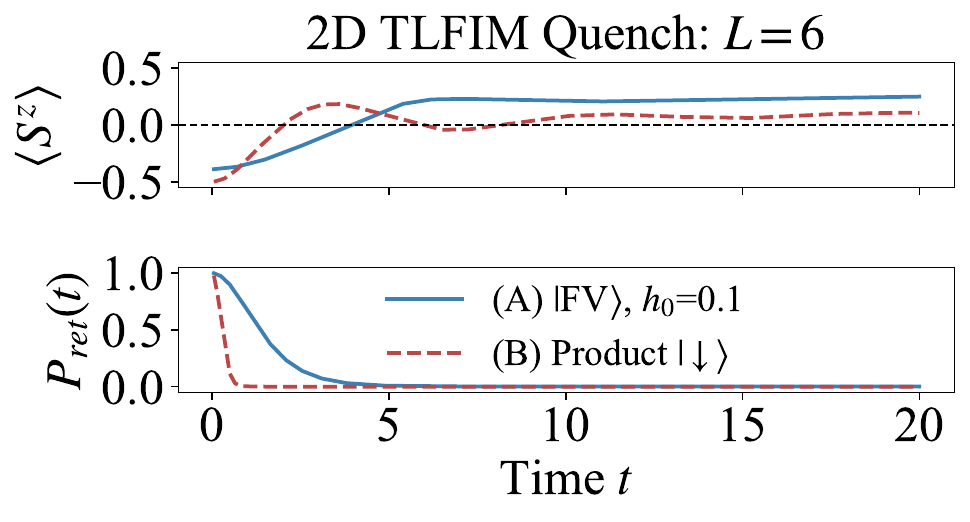}
        \label{fig:tlfim_quench_6by6}
    \end{subfigure}
    \begin{subfigure}[t]{0.48\linewidth}
        \centering
        \caption{}
        \includegraphics[width=\linewidth]{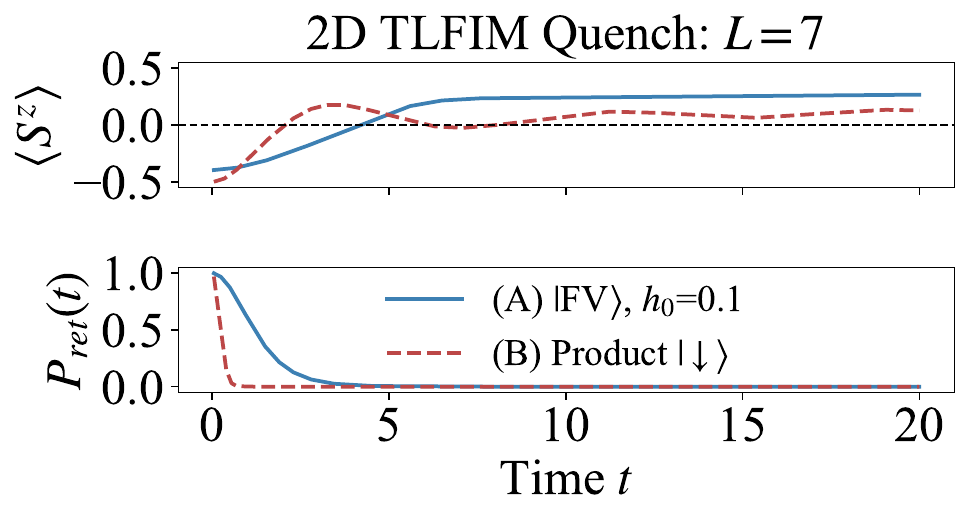}
        \label{fig:tlfim_quench_7by7}
    \end{subfigure}

    \caption{Quench dynamics of the 2D TLFIM for system sizes (a) $L=4$, (b) $L=5$, (c) $L=6$, and (d) $L=7$. We set $J=-1.0$, $g=1.0$, post-quench field $h_q=-0.2$, initial symmetry-breaking field $h_0=0.1$, and bond dimension $\chi=256$. In each panel, the top plot shows the mean longitudinal magnetisation $\langle S^z\rangle(t)$, and the bottom plot shows the return probability $P_{\mathrm{ret}}(t)$. The blue line corresponds to the quench starting from the $\mathbb{Z}_2$-broken correlated DMRG ground state of $H_0$, while the red dashed line shows the quench starting from the product state $\ket{\psi_{0}}$.}
    \label{fig:tlfim_quench_size_compare}
\end{figure*}

MPS simulations are performed here in which the 2D lattice is mapped to a 1D chain using a snake ordering, which preserves the square-lattice connectivity while mapping vertical nearest-neighbour bonds to longer-range interactions along the chain \cite{stoudenmire2012studying}. Since our goal is to study the role of initial-state entanglement in FV decay, we consider the non-equilibrium dynamics of Eq. (\ref{eq:Ising}) following a quench and numerically examine a range of scenarios across different lattice sizes, degrees of entanglement, and dimensionalities. For these calculations, the many-body wavefunction is represented as an MPS and evolved using the two-site time-dependent variational principle (TDVP)~\cite{haegeman2011time,PhysRevB.94.165116}. 
Truncation is controlled by the bond-dimension caps $\chi_{\mathrm{DMRG}}$ for state preparation and $\chi_{\mathrm{q}}$ for time evolution, together with the singular-value cutoff $\texttt{svd\_min}$. 
Adaptive bond dimensions up to $\chi=256$ are used to verify the convergence of local and global observables.
MPS is used for the smaller lattices and the comparisons in Fig.~\ref{fig:max_cluster_size_dynamics}, whereas a binary TTN with one-site TDVP is employed for the larger systems in Figs. \ref{fig:TTN_large_cluster_dynamics} and \ref{fig:TTN_edge_var}.

Fig.~\ref{fig:tlfim_quench_size_compare} reveals a striking difference between $\ket{\psi_0}$ initialisation and the correlated DMRG ground state of $H_0$. For $\ket{\psi_0}$, we can see that $P_{\mathrm{ret}}(t)$ decays sharply, indicating rapid dephasing across all system sizes. In contrast, the correlated state retains finite $P_{\mathrm{ret}}(t)$ for significantly longer, with more pronounced oscillations for smaller lattices [see Fig.~\ref{fig:tlfim_quench_size_compare}(a,b)] and delayed decay that signals the presence of a long-lived metastable regime. We observe that this behaviour persists even for larger systems, Fig.~\ref{fig:tlfim_quench_size_compare}(c,d). Although here finite-size effects are more apparent, i.e., faster decay and suppressed oscillation amplitudes, the correlated state consistently shows slower loss of $P_{\mathrm{ret}}(t)$. In terms of $M_z(t)$, both states begin from a FV polarisation configuration. Here, for $\ket{\psi_0}$ we observe sharper initial growth that indicates swift decay via uncorrelated local spins that act as nucleation seeds for the formation of TV bubbles. Notably, the magnetisation trajectories converge at late times only for the smallest $L=4$ lattice, which is consistent with initial-state correlations affecting the dynamics over longer times in the larger systems considered.

\begin{figure*}[ht!]
    \centering
    \includegraphics[width=1.0\linewidth]{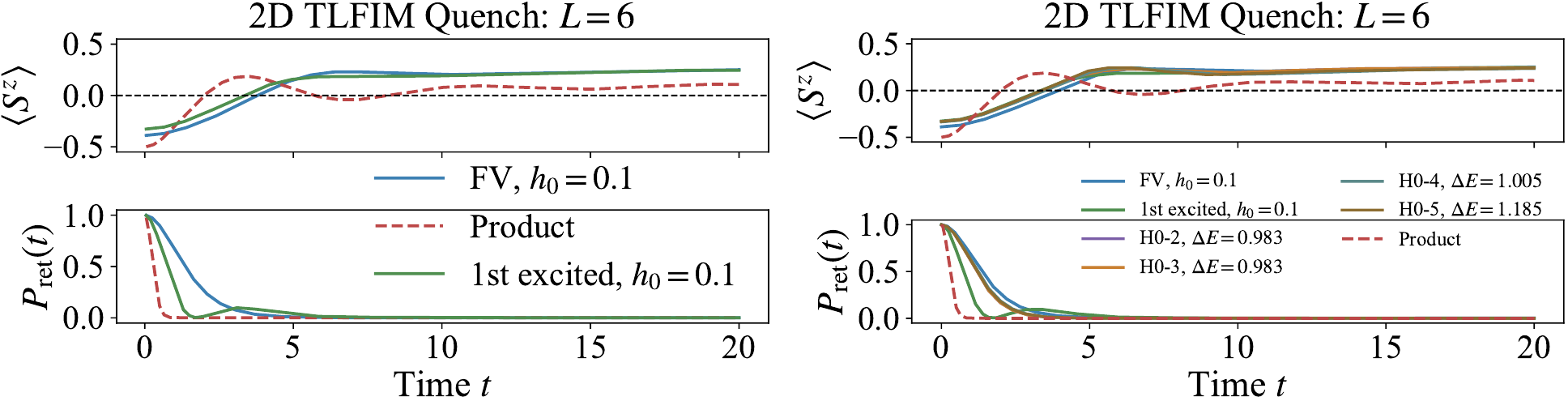}
    \caption{
    Evolution from different initial states.
    \textbf{Left:} Quench dynamics of the 2D TLFIM on an $L=6$ lattice starting from (i) the correlated DMRG ground state of $H_0$, (ii) the product state $\ket{\psi_0}$, and (iii) the first excited state of $H_0$.
    \textbf{Right:} Corresponding dynamics for additional eigenstates of the pre-quench Hamiltonian.
    These mutually orthogonal eigenstates exhibit enhanced metastability relative to the product state following the longitudinal-field quench.
    The remaining parameters are the same as in Fig.~\ref{fig:tlfim_quench_size_compare}.
    }
    \label{fig:excited_state_evolution}
\end{figure*}
To probe the role of correlations beyond product-state initialisation, in Fig.~\ref{fig:excited_state_evolution} we compare the quench dynamics of the correlated DMRG ground state with that of the first excited state of $H_0$ and different eigenstates, which can help us determine whether intrinsic excitations in the FV act as nucleation seeds that accelerate the decay. Evidently, the ground and first excited states exhibit distinct initial magnetisations, reflecting their different internal correlation structures. However, their long-time magnetisation values nearly converge, indicating that the asymptotic state is largely determined by $H_{\mathrm{q}}$. The transient dynamics, in contrast, depend strongly on the initial-state structure. The observed decay of $P_{\mathrm{ret}}(t)$ demonstrates a dynamical stability hierarchy associated with the degree of initial-state correlations. While the first excited state already contains nontrivial correlations relative to $\ket{\psi_{0}}$, it exhibits faster loss of coherence than the DMRG ground state, indicating that low-energy excitations only weaken (but do not eliminate) the stabilising effect of initial entanglement. These observations further support the view that metastable behaviour is controlled not simply by the energy above the ground state of $H_0$, but by the structure of correlations in the initial state.

\subsubsection{Supplemental simulations and numerical details for Figs.~3 and~4}
\label{sec:SMb2}

For the large-scale 2D results shown in Figs. \ref{fig:TTN_large_cluster_dynamics} and \ref{fig:TTN_edge_var} of the main text, the real-time dynamics was simulated using the TN backend of \texttt{QTea} \cite{qtealeaves}. To access larger 2D lattices, we employed a binary TTN ansatz rather than a snake-ordered MPS. In the TTN representation, the physical spins are recursively grouped into a tree, and the entanglement capacity across each tree bond is controlled by the maximum bond dimension \(\chi\). This structure is advantageous for 2D systems~\cite{krinitsin2025time,pavevsic2025scattering}: even when the physical sites are arranged in the same snake order, the TTN ansatz connects distant regions through a hierarchical tree rather than through a single 1D chain, thereby providing a more efficient representation of long-range correlations than a snake-ordered MPS.

\begin{figure}
    \centering
    \begin{subfigure}[t]{0.48\linewidth}
        \centering
        \includegraphics[width=\linewidth]{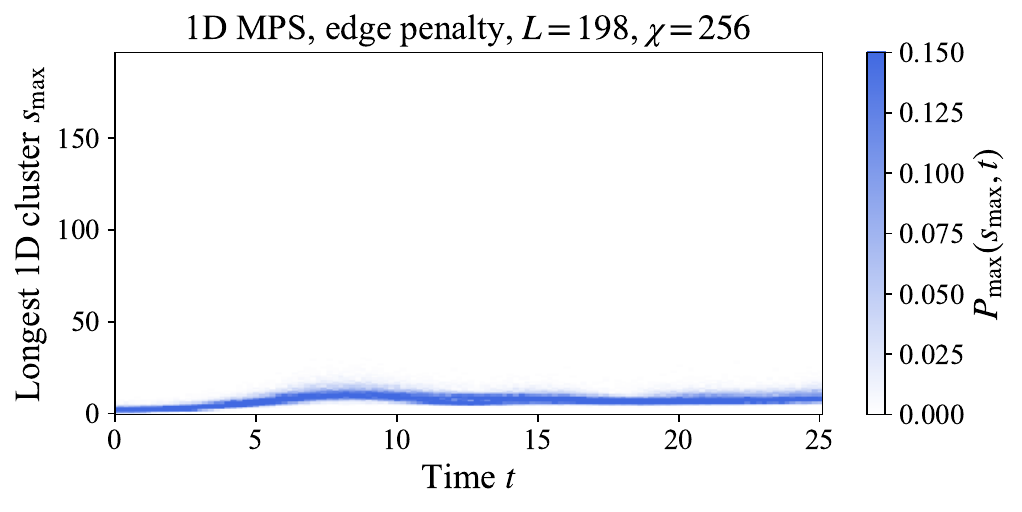}
        \label{fig:1d_edge_pin_fv_L198_g_0.3_a}
    \end{subfigure}
    \hfill
    \begin{subfigure}[t]{0.44\linewidth}
        \centering
        \includegraphics[width=\linewidth]{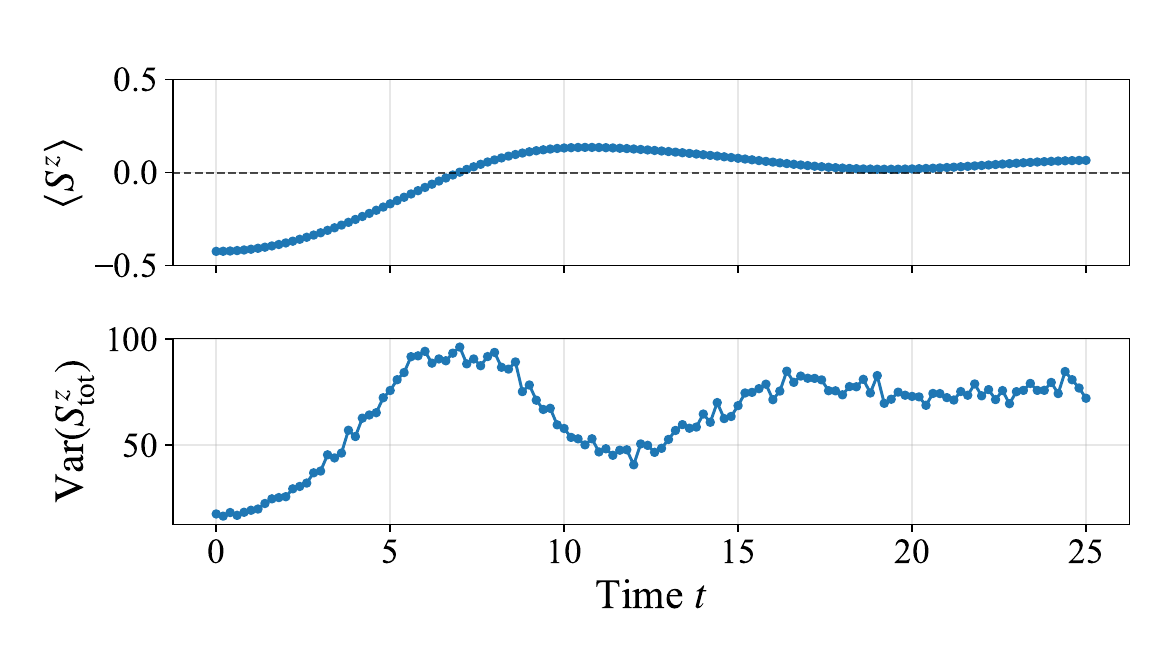}
        \label{fig:1d_edge_pin_fv_L198_g_0.3_b}
    \end{subfigure}
    \caption{
    Edge-pinned 1D FV dynamics with the same number of active spins as the edge-pinned 2D simulation in Fig. \ref{fig:TTN_edge_var}.
    The system is a chain of length \(L=198\), containing \(196\) active spins, initialised in the FV and evolved using an MPS with bond dimension \(\chi=256\).
    The variance \(\mathrm{Var}(S^z_{\mathrm{tot}})\) is more than one order of magnitude smaller than in the \(L\times L\) 2D FV decay model shown in Fig. \ref{fig:TTN_edge_var}, and the two-branch structure in \(P_{\max}(s_{\max},t)\) is absent. In this 1D simulation, the transverse field is set to \(g=0.3\), so that the 1D model lies in its ferromagnetically ordered regime.
    }
    \label{fig:1d_edge_pin_fv_L198}
\end{figure}

As a 1D benchmark, Fig.~\ref{fig:1d_edge_pin_fv_L198} shows edge-pinned FV dynamics in a chain with the same number of active spins as the 2D calculation in Fig. \ref{fig:TTN_edge_var}. Although the boundary nucleation channel is suppressed by endpoint pinning, the resulting \(\mathrm{Var}(S^z_{\mathrm{tot}})\) is more than one order of magnitude smaller than in the 2D edge-pinned case, and the same macroscopic two-branch structure is not observed. For the 2D simulations, the initial $\ket{\rm{FV}}$ was prepared by a static variational ground-state optimisation in the TTN manifold, with the initial and minimum TTN bond dimensions set equal to the requested \(\chi\) in order to avoid trapping the optimisation in an undersized tree manifold. The subsequent quench dynamics was evolved with TDVP: one-site TDVP was used for the TTN calculations, while the MPS benchmark was evolved using two-site TDVP. For the 2D calculations, projective measurements in the computational basis were performed during the TDVP evolution and mapped back to the snake-ordered square lattice to extract cluster observables. The simulations ran in double-precision complex arithmetic using the CUDA/CuPy GPU backend; in production runs this backend was assigned to an NVIDIA H100 GPU.

\begin{figure}[t!]
    \centering
    \begin{subfigure}[t]{0.48\linewidth}
        \centering
        \includegraphics[width=\linewidth]{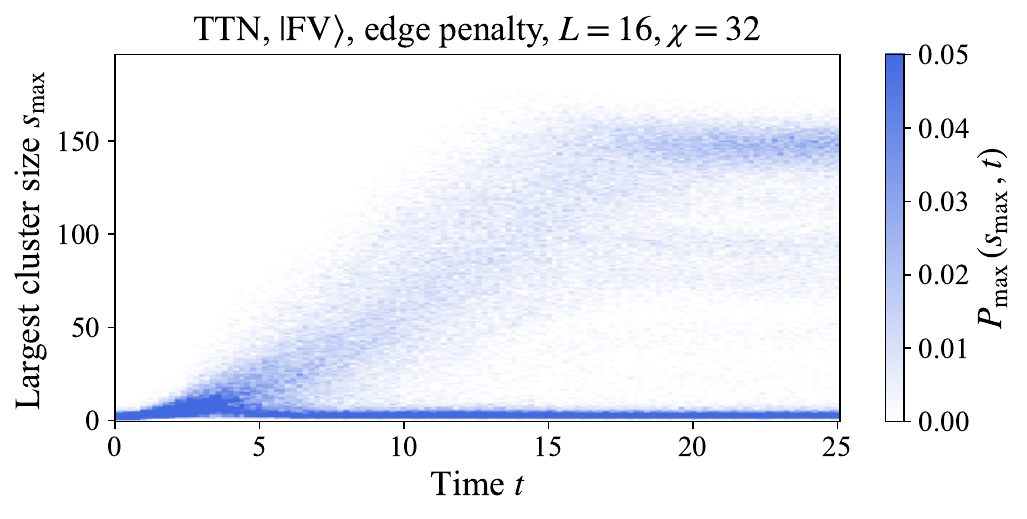}
        \label{fig:FV_edge_pin_Chi_32}
    \end{subfigure}
    \hfill
    \begin{subfigure}[t]{0.48\linewidth}
        \centering
        \includegraphics[width=\linewidth]{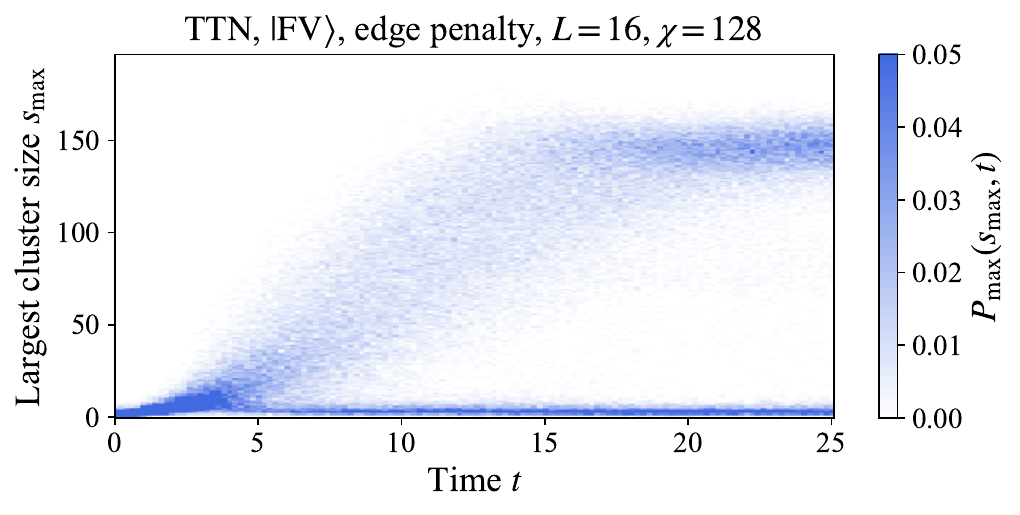}
        \label{fig:FV_edge_pin_Chi_128}
    \end{subfigure}
    \caption{
    Convergence check of the static optimisation followed by one-site TDVP as the TTN bond dimension \(\chi\) is increased.
    We show the dynamics of \(P_{\max}(s_{\max},t)\) starting from the FV under the edge-pinned Hamiltonian with \(h_{\mathrm{edge}}=20\).
    At each time step, the wavefunction is projectively sampled \(800\) times.
    The one-site TDVP time step is \(dt=0.02\).
    The \(\chi=32\) simulation already captures the same initial branching and late-time distribution features as the \(\chi=128\) simulation.
    }
    \label{fig:FV_edge_pin_benchmark}
\end{figure}

\begin{figure}[t!]
    \centering
    \begin{subfigure}[t]{0.48\linewidth}
        \centering
        \includegraphics[width=\linewidth]{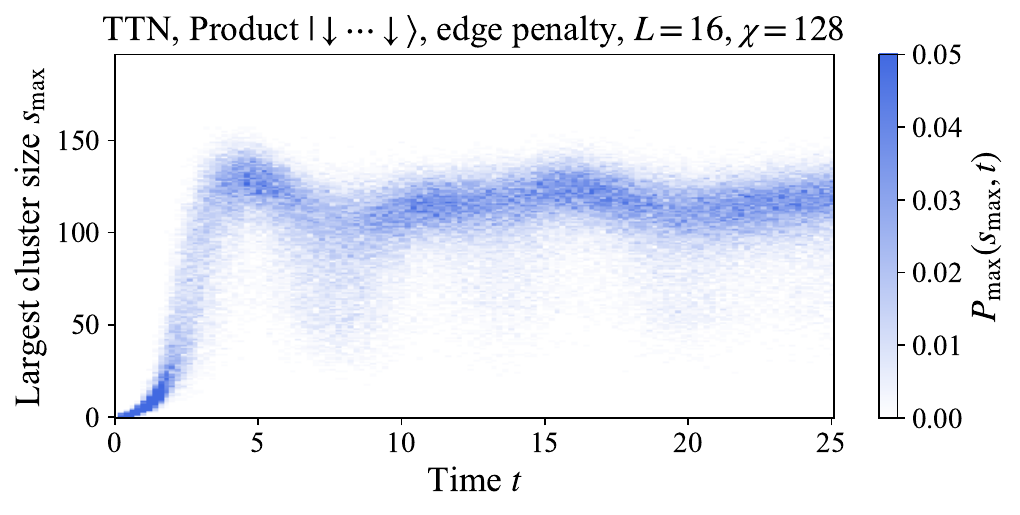}
        \label{fig:alldown_edge_pin_L16_Chi128}
    \end{subfigure}
    \hfill
    \begin{subfigure}[t]{0.48\linewidth}
        \centering
        \includegraphics[width=\linewidth]{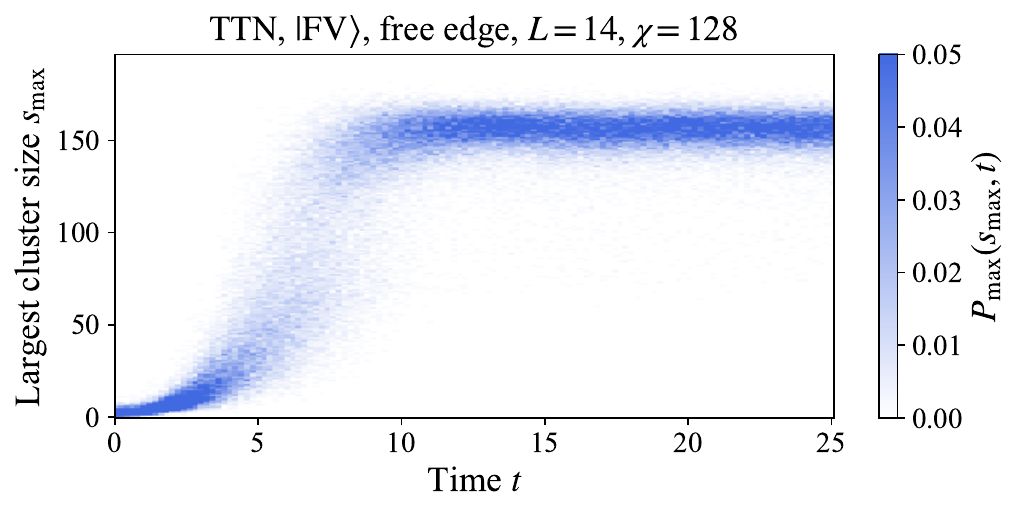}
        \label{fig:FV_edge_free_L14_Chi128}
    \end{subfigure}
\caption{
Control simulations testing the conditions required for the macroscopic two-branch structure observed in Fig. \ref{fig:TTN_edge_var}.
\textbf{Left:} Dynamics initialised from $\ket{\psi_{0}}$ under the edge-pinned Hamiltonian.
\textbf{Right:} Dynamics initialised from $\ket{\rm{FV}}$ without edge pinning.
In the edge-pinned case, \(L=16\) and the number of active spins is \((L-2)^2\), with edge-pinning field \(h_{\mathrm{edge}}=20\).
In the unpinned case, \(L=14\).
For both simulations, the one-site TDVP time step is \(dt=0.02\) and the TTN bond dimension is \(\chi=128\).
Together with the 1D benchmark in Fig.~\ref{fig:1d_edge_pin_fv_L198}, these controls indicate that FV initialisation, edge pinning, and 2D connectivity are all needed for the macroscopic two-branch structure.
}
    \label{fig:all_down_edge}
\end{figure}

Fig.~\ref{fig:FV_edge_pin_benchmark} shows a bond-dimension convergence check for the edge-pinned 2D FV dynamics. The \(\chi=32\) simulation already captures the same initial branching and late-time distribution features as the \(\chi=128\) simulation. Fig.~\ref{fig:all_down_edge} presents two control simulations. Together with the 1D benchmark in Fig.~\ref{fig:1d_edge_pin_fv_L198}, these results show that the cat-state generation in Fig. \ref{fig:TTN_edge_var} requires FV initialisation, edge pinning, and 2D connectivity. Starting from the all-spin-down state with edge pinning, starting from $\ket{\rm{FV}}$ without edge pinning, or using a genuine 1D chain geometry does not produce the same macroscopic two-branch structure.

\end{document}